\documentclass[letter]{article}

\usepackage[top=2.8cm, bottom=2.8cm, left=2.4cm, right=2.4cm]{geometry}
\usepackage[utf8]{inputenc}  
\usepackage[T1]{fontenc}     
\usepackage{amsmath,amssymb,lmodern} 
\usepackage{graphicx}
\usepackage{hyperref}
\usepackage{color}
\usepackage{multicol} 

\newcommand{\be}{\begin{equation}}
\newcommand{\ee}{\end{equation}}
\newcommand{\bea}{\begin{eqnarray}}
\newcommand{\eea}{\end{eqnarray}}
\newcommand{\bal}{\begin{eqnarray}}
\newcommand{\eal}{\end{eqnarray}}

\newcommand{\no}{\nonumber}

\begin{document}

\title{\begin{flushright}
\normalsize PI/UAN-2019-665FT
\end{flushright}
\vspace{5mm}
{\bf de Sitter symmetries and inflationary scalar-vector models}}

\author{
\hspace{-8mm}\textbf{Juan P. Beltr\'an Almeida$^{1,}$\thanks{e-mail: \texttt{juanpbeltran@uan.edu.co}}, Josu\'e Motoa-Manzano$^{2,}$\thanks{e-mail:
\texttt{josue.motoa@correounivalle.edu.co}}, C\'esar A. Valenzuela-Toledo$^{2,}$\thanks{e-mail: \texttt{cesar.valenzuela@correounivalle.edu.co} }} \\ \\
\hspace{-12mm}\textit{$^1$Departamento de F\'isica, Facultad de Ciencias, Universidad Antonio Nari\~no,}\\
\hspace{-12mm}\textit{Cra 3 Este \# 47A-15, Bogot\'a D.C., Colombia}\\
\hspace{-12mm}\textit{and} \\
\hspace{-12mm}\textit{$^2$Departamento de F\'isica, Universidad del Valle,}  \\
\hspace{-12mm}\textit{Ciudad Universitaria Mel\'endez, Santiago de Cali 760032, Colombia.} \\
}

\maketitle

\begin{abstract}
In this paper we study the correspondence between a field theory in de Sitter space in $D$-dimensions and a dual conformal field theory in a euclidean space in $(D-1)$-dimensions. In particular, we investigate the form in which this correspondence is established for a system of interacting scalar and a vector fields propagating in de Sitter space. We analyze some necessary (but not sufficient) conditions for which conformal symmetry is preserved in the dual theory in $(D-1)$-dimensions, making  possible the establishment of the correspondence. The discussion that we address on this paper is framed on the context of {\it inflationary cosmology}, so, the results obtained here pose some relevant possibilities of application to the calculation of the fields's correlation functions and of the {\it primordial curvature perturbation} $\zeta$,  in inflationary models including coupled scalar and vector fields. \\ \\
\vspace{0.5cm}
\end{abstract}
\begin{multicols}{2}
\section{Introduction}
\label{intro}
The study of the symmetries 
is very important when we look for fundamental descriptive features of a particular physical system and solve its dynamics, for example, it is well known that through the Noether theorem we can identify conserved quantities related to the symmetries, 
which allows us to discover intrinsic fundamental characteristics that will be reflected on the dynamics of the system. It is also well 
known that the geometry of the {\it inflationary period} of our universe is described with great accuracy by the de Sitter space.
This spacetime is similar to the Minkowski spacetime in the sense that it is a maximally symmetric space, which means that in $D$ dimensions, the number of symmetry generators is $D(D+1)/2$. This fact defines and restricts the form of the kinematic features and the evolution of a system on this space. During the inflationary period the symmetries of the de Sitter group play an essential role on the description of the primordial perturbations and their statistical properties.  
In particular, by using the correspondence between the de Sitter symmetry group in $D$-dimensions and the conformal group in an euclidean $(D-1)$-dimensional ``boundary'' space  \cite{Strominger:2001pn}, 
it is possible to calculate general important features of the system, such as the correlations functions of the fields that govern the dynamics of the inflationary Universe \cite{riotto}, which are necessary to calculate the statistical descriptors of the primordial curvature perturbation by using, for example, the $\delta {\rm N}$ formalism \cite{Sasaki95,Nakamura96,Starobinsky86,Lyth04dn,Dimopoulos08b} or cosmological perturbation theory \cite{Gumrukcuoglu10,Bartolo12} . To do that, we need first to calculate one of the fundamental quantities of the conformal field theory description: the {\it conformal weights} of the fields.

The correspondence between a field theory in de Sitter space and a dual theory in a euclidean boundary space, one dimension less than the de Sitter space, is what we shall refer to as dS/CFT correspondence and it is at the core of the approach that we are pursuing here. This correspondence was proposed initially for free non interacting single fields propagating on de Sitter space \cite{Bousso:2001mw} but some considerations about the interacting case was discussed in the literature \cite{Spradlin:2001nb}.  
In this paper, we follow the approach proposed by \cite{riotto} to study the case of a coupled scalar-vector system. In the context that we are framing our discussion, the scalar field is responsible for the inflationary expansion and the vector field is partly responsible of the generation of the primordial curvature perturbation $\zeta$. The inclusion of vector fields on the inflationary dynamics permits the study of several interesting possibilities such as the presence of statistical anisotropies, parity violating patterns, the origin of primordial magnetic fields, etc. (see for instance \cite{Caprini:2014mja,Maleknejad12,Soda12,Dima10a} and references therein). For concreteness we will settle the discussion in four dimensional de Sitter space but the results obtained can be generalised straightforwardly to any dimensions.  

This paper is organized as follows. In section II we review the symmetries and some basics of de Sitter space. Next, on sections III and IV we shall describe the transformation laws imposed by these symmetries over the conformal fields on the dual theory and we calculate the associated conformal weights of those fields. In section V we briefly review the results for the scalar and the vector field separately and then, in section VI we consider the coupled  scalar-vector system. Finally, in section VII we present our conclusions and discuss some possible applications of the results obtained.    
\section{Symmetries of the de Sitter Space}
The de Sitter spacetime can be described in several useful coordinate systems (for reviews see for instance \cite{Spradlin:2001pw, Leblond:2002ns,  Leblond2}), but for our interests, we will chose the 4D conformal planar coordinates in which the line element is written as 
\be
\label{ds}
ds^2=\frac{1}{\left(H\tau\right)^2}\left(-d\tau^2+d\vec{x}^2\right),
\ee
where $H$ is the Hubble parameter and  $\tau$ is the {\it conformal time}. In this coordinate system it is easy to see that the line element (\ref{ds}) is invariant under spacial translations and rotations on $\tau = constant$ sections:
\be
 x'_i=a_i+R_{ij}x_j,\label{ltz}
\ee
where $a_i$ is a 3-dimensional constant vector and  $R_{ij}$ is an $O(3)$ matrix  satisfying the condition $R_{ik}R^{kj}=\delta_i^j$ and representing a three dimensional rotation. We have a total of six symmetry generators included in the transformation (\ref{ltz}) and they affect only the spacial sections of the spacetime. Notice also that we allow the possibility of spacial reflexion  $x^i \rightarrow -x^i $ in this group of transformations. Spacial reflexion is an explicit symmetry of the line element (\ref{ds}) and we will pay attention to this specific transformation, when we consider the possibility of introducing parity violating models in the presence of vector fields. There are two additional transformation which mixes the time and the spacial coordinates, the spacetime dilatation:  
\be 
 x'^\mu=\lambda x^\mu \rightarrow \quad x'_i=\lambda x_i, \qquad \tau\rightarrow \tau'=\lambda \tau,\label{dilation}
\ee
where $\lambda$ is a constant factor 
and the special conformal transformation:
\be\label{CST}
x'^\mu=\frac{x^\mu+b^\mu  {x}^2 }{1+2\vec{b}\cdot\vec{x}+
b^2  {x}^2 },
\ee
or
\be\no
x'_i=\frac{x_i+b_i\left(-\tau^2+\vec{x}^2\right)}{1+2\vec{b}\cdot\vec{x}+
b^2\left(-\tau^2+\vec{x}^2\right)},
\ee
and
\be\no
 \tau'=\frac{\tau}{1+2\vec{b}\cdot \vec{x}+b^2\left(-\tau^2+\vec{x}^2\right)},
\ee
with $b^\mu = (0, b_i)$ a three dimensional vector which generates the transformation. 
In the last expressions we use the notation $x^2 = \eta_{\mu\nu} x^\mu x^{\nu}$, $\eta_{\mu\nu}= {\rm diag}(-1, 1, 1,1)$, $\vec{x}^2 = \delta_{ij} x^ix^j$ and $\vec{x}\cdot\vec{b} = \delta_{ij} b^ix^j$. Together, the transformations (\ref{ltz}), (\ref{dilation}) and (\ref{CST}) tell us that  de Sitter space has ten symmetry generators, so, this is a maximally symmetric spacetime.  It can be seen that (\ref{CST}) is constructed out by the composition of three consecutive transformations, namely, an inversion, written as
\be
x'^\mu=\frac{x^\mu}{x^2},
\ee
or
\be
 x'_i=\frac{x_i}{-\tau^2+\vec{x}^2}, \quad
 \tau'=\frac{\tau}{-\tau^2+\vec{x}^2},\label{inv}
\ee
then a translation ($\tau'\rightarrow\tau''=\tau'$ and $x'_i\rightarrow x''_i=x'_i+b_i$) and then an inversion again. Actually, the special conformal transformation (\ref{CST}) can be expressed as:
\be \label{invsct}
\frac{ {x'}^\mu}{x'^2} = \frac{x^\mu}{x^2} + b^\mu.
\ee
It becomes useful in the following to have the Jacobian matrices associated to the coordinate transformations of the de Sitter isometries. Particularly, we will need the Jacobian  for the  dilatation and for special conformal transformation (\ref{dilation}) and (\ref{CST}). Nevertheless, as we said before, the special conformal transformation can be obtained by composing translations and inversions, then, we use the transformation (\ref{inv}) instead of (\ref{CST}). 
The Jacobian matrix of the inversion, can be derived directly for (\ref{inv}) obtaining  
\be\label{invdev}
\frac{\partial x'^\mu}{\partial x^\nu}=\frac{1}{x^2}\left(\delta_\mu^\nu-2 \frac{x^\mu \eta_{\nu \alpha} x^\alpha}{x^2}\right)=\frac{1}{x^2}J_\nu^\mu,
\ee
where  $J_\mu^\nu$ satisfies the orthogonality  relation $J_\mu^\sigma J_\sigma^\nu=\delta_\mu^\nu$. The associated Jacobian determinant of the above transformation is
\be \label{djacinv}
\det\left(\frac{\partial x'^\mu}{\partial x^\nu}\right)=-\frac{1}{\left(x^2\right)^4}. 
\ee 
Using (\ref{inv}) we can also calculate the Jacobian of the inverse transformation $x'\rightarrow x$, obtaining 
\be
\frac{\partial x^\mu}{\partial x'^\nu}=x^2 J_\nu^\mu, \quad \det\left(\frac{\partial x^\mu}{\partial x'^\nu}\right)=\left(x^2\right)^4.
\ee
On the other hand, for the dilatations we have:
\begin{align}
x'^\mu&=\lambda x^\mu\:,\\ \label{djacdil}
\frac{\partial x'^\mu}{\partial x^\nu}&=\lambda \delta_\nu^\mu, \quad 
\det\left(\frac{\partial x'^\mu}{\partial x^\nu}\right)=\lambda^4,\\
\frac{\partial x^\mu}{\partial x'^\nu}&=\frac{1}{\lambda}\delta_\nu^\mu\; \quad \det\left(\frac{\partial x^\mu}{\partial x'^\nu}\right)=\lambda^{-4}.
\end{align}
Then, with the previous results we can also calculate the change in the volume element, which reads:
\be
d\tau' d^3x'=\left|\det\left(\frac{\partial x'^\mu}{\partial x^\nu}\right)\right|d\tau d^3x.
\ee

On next sections we will exploit the correspondence between the de Sitter isometry group represented by the transformations below  and the conformal group in $\mathbb{R}^3$. In the essence of this correspondence lies the idea that de Sitter isometries act as the conformal group transformations in $\mathbb{R}^3$, a fact that we will show on next section. Particularly, we will consider the asymptotic region, where inflationary perturbations are frozen, after horizon crossing at supper Hubble scales which in these coordinates happens for $-\tau \ll |\vec{x}|$. Then, in the following section we will recall some basis of the conformal group and introduce the relevant terminology that will allow us to construct a conformal field theory representation of the fields in de Sitter space. An important part of the following discussion is devoted to the {\it conformal weight} of the fields in the conformal field theory representation. The conformal weight of the field  is a crucial element for expressing the transformation rules of the fields and the symmetries of their correlation functions which are the meaningful objects constructed in the theory because they encode the statistical  properties of the theory.   

\section{Conformal group basics and relation with de Sitter group  symmetries  
}
In an Euclidean space, the conformal group is defined as the set of coordinate transformations $x\rightarrow x'$ that leave invariant the angles between two vectors in this space, or, equivalently, the group of transformations that leave invariant the metric up to a factor, this is
\be \label{confg}
g_{\mu \nu}(x) \rightarrow  g'_{\mu \nu}(x') = \omega(x)g_{\mu \nu}(x),
\ee
where $\omega(x)$ is an arbitrary function of the coordinates. It is easy to see that the transformations (\ref{ltz}), (\ref{dilation}) and  (\ref{CST})  act as conformal transformations on $\mathbb{R}^3$ for the euclidean metric, i.e. $g_{ij} = \delta_{ij} $, in (\ref{confg}) with the corresponding conformal factor $\omega$:
\be \label{tr3d}
x'_i = a_i+R_{ij} x_j \quad  \rightarrow \quad \omega = 1,
\ee
\be \label{dil3d}
x'_i = \lambda x_i \quad  \rightarrow \quad \omega = \lambda^{-2}, 
\ee
\bea
x'_i &=&\frac{x_i+b_i \vec{x}^2 }{1+2\vec{b}\cdot\vec{x}+
b^2\left(\vec{x}^2\right)}  \no\\\label{sct3d}
& \rightarrow & \omega = (1+2\vec{b} \cdot \vec{x} + b^2 \vec{x}^2 )^2.
\eea
Then, we can see explicitly that the de Sitter symmetry group in four dimensions induce a conformal group in the three dimensional space. In few words, we can interpret the symmetries (\ref{tr3d}) to (\ref{sct3d}) as the {\it asymptotic symmetry group} of the  boundary region of the four dimensional de Sitter space time. In our context, the asymptotic region is located at super horizon scales when the inflationary perturbations evolve classically and carry the information of the physical mechanism that drives the inflationary expansion. In the coordinates employed here, and in the previous expressions, we have used that the super horizon limit  which constitutes our asymptotic euclidean space lies in the region $|\tau| \ll |\vec{x}|$.  It its worth to notice that we obtain ten symmetry group generators as we expect for the conformal group acting on three dimensional space. 
 \\
Additionally, or equivalently to the expression (\ref{sct3d})  we can write the inversion transformation and its associated conformal factor as:
\be\label{inv3d}
x'_i =\frac{x_i  }{\left(\vec{x}^2\right)} \;  \rightarrow \; \omega = (\vec{x}^2 )^2.
\ee
So far we were talking about the symmetries and the geometry of the spacetime, and now we turn to the fields on the theory. The relevant objects that we need for the conformal field theory description, are the ones that we shall call a {\it  primary field}. A  {\it  primary field} is an object which  transforms according to the following rule  under conformal transformations
\bal\label{tl}\no
T_{i_1...i_n}(x) \rightarrow T'_{i_1...i_n}(x')=\left|\det\left(\frac{\partial x'^l}{\partial x^k}\right)\right|^{\frac{n-\Delta_T}{d}}\times&\\
\frac{\partial x^{j_1}}{\partial x'^{i_1}}...\frac{\partial x^{j_n}}{\partial x'^{i_n}}T_{j_1...j_n}(x),&
\eal

where $n$ is the order of the tensor, $d$ is the dimension of the spacetime and $\Delta_T$ is the {\it conformal weight} of the field  \cite{riotto}. When dealing with the conformal dual theory to the theory in de Sitter space, we will consider the Euclidean space $\mathbb{R}^3$.   
By using the above transformation, we can find the conformal dimension $\Delta_T$ of different fields that are 
involved in the theory. It is beyond the scope of this short article to enter in the details of the conformal field theory machinery, we will just restrict to the introduction of the fundamental elements necessary for our purpose. A complete study of conformal field theory can be found for instance in \cite{Ginsparg:1988ui}. \\

\section{Fields on de Sitter space}
Now we study of fields propagating on de Sitter space. With the results obtained in the previous section we can evaluate the conformal weight, which is an essential quantity necessary for the conformal field theory description of the fields. We are interested in the coupled scalar vector system, but, before that, and mainly for illustrative reasons and for developing the technique introducing the necessary elements, we study the single scalar field and the single vector field cases separately. After that and invoking the results from the single field case, we will face the coupled system. 
\subsection{Single scalar field.}
The action for a single massive scalar field propagating on de Sitter space in the coordinates (\ref{ds}) is given by
{\small\bal\no
S_\phi&=-\frac{1}{2}\int d^4x\sqrt{-g}\left\{\partial_\mu \phi \partial ^\mu\phi+m_\phi^2\phi^2\right\} \\
&=-\frac{1}{2}\int \frac{d\tau d^3x}{ H^2\tau^2}\left\{\eta^{\mu \nu}\partial_\mu \phi \partial_\nu\phi+\frac{m_\phi^2\phi^2}{H^2\tau^2}\right\},\label{aces1}
\eal}
where $m_{\phi}$ is the mass of the field. Now, we study the symmetries of the theory in order to learn about the transformation of the fields. To emphasise the difference between the properties of the scalar field in de Sitter space and in $\mathbb{R}^3$, let us suppose that we write the transformation law for a scalar in de Sitter as in (\ref{tl}):
\be
\phi'=\left|\det\left(\frac{\partial x'^\mu}{\partial x^\nu}\right)\right|^{-\frac{\Delta_\phi}{4}}\phi,
\ee 
and for the action
{\small \be\no
S'_\phi=-\frac{1}{2}\int d\tau' dx'^3\frac{1}{H^2\tau'^2}\left\{\eta^{\mu \nu}\partial'_\mu \phi' \partial'_\nu\phi'
+\frac{m_\phi^2\phi'^2}{H^2\tau'^2}\right\}.
\ee}

Under the inversion (\ref{inv}) we have: $\phi'=(x^2)^{\Delta_\phi}\phi\;$ and 
\bea\no
S'_\phi&=&\int d\tau dx^3\frac{1}{H^2\tau^2}\Big\{-4\Delta^2_\phi(x^2)^{2\Delta_\phi}\phi^2\\\no
&&-4\Delta_\phi(x^2)^{2\Delta_\phi-1}\phi x^\mu \partial_\mu \phi-(x^2)^{2\Delta_\phi}\\
&&\times\big[\eta^{\mu \nu}\partial_\mu \phi \partial_\nu\phi+\frac{m_\phi^2\phi^2}{H^2\tau^2}\big]\Big\}\,,\no
\eea
where we have used $\partial'_\mu=x^2 J_\mu^\nu \partial_\nu$. 
Demanding invariance of the action $S_\phi=S'_\phi$ we see that it is necessary to set $\Delta_\phi=0$, and then $\phi' = \phi$. It is easy to arrive to the same result if we apply the dilatation (\ref{dilation}) and it is trivial for the translations and three dimensional rotations (\ref{ltz}) because the  transformation  matrix is orthogonal. Certainly, this result does not come as a surprise at all, since we are talking about of a scalar field.  Nevertheless it is worth to stress this fact because we are going to see that asymptotically, in the super horizon limit the dual field associated to the scalar transform according with (\ref{tl}) with a conformal weight different than $\Delta_\phi=0$. 

Now, we study the behaviour of the associated conformal field of the scalar filed in the three dimensional space $\mathbb{R}^3$. To this end it is necessary to solve the equation of motion and go to super horizon scales. The equation of motion derived from (\ref{aces1}) is
\be\label{eqphi}
 \ddot{\phi}-\frac{2}{\tau}\dot{\phi}-\nabla^2 \phi+\frac{m_\phi^2\phi}{H^2\tau^2}=0.
\ee
To solve this equation we go to momentum space. Using the Fourier transform of the field, that is  $F(\tau, \vec{x}) = \int \frac{d^{3} x}{ (2\pi )^{3/2}} \tilde{F}(\tau, \vec{k}) {\it e}^{i \vec{k} \cdot \vec{x}}$, the resulting equation is:
\be
\ddot{\tilde{\phi}}-\frac{2}{\tau}\dot{\tilde{\phi}} + k^2 \tilde{\phi} + \frac{m_\phi^2 \tilde{\phi} }{H^2\tau^2}=0,
\ee
which can be fully solved analytically in terms of Bessel functions $J_\nu$ and $Y_\nu$:
\be\label{solphi}
\tilde{\phi}(\tau,\vec{k})=\tau^{3/2}\left(C_1(\vec{k}) J_{h}(k\tau)+C_2(\vec{k}) Y_{h}(k\tau)\right),
\ee
where $C_1$ and $C_2$ are constants which depends on boundary initial conditions and 
\be\label{h} h=\sqrt{\frac{9}{4}-\frac{m_\phi^2}{H^2}}. \ee
Now, we need the  behaviour of the solution (\ref{solphi}) at super horizon scales, which in momentum space are achieved when $-k\tau \ll 1$ \cite{PhysRevD.90.064010}. By expanding the Bessel functions in that regime and going back to $(\tau, \vec{x})$ space, we arrive to 
\be
{\phi} (\tau, \vec{x} ) \approx \tau^{\frac{3}{2}+h}\sigma_{+}(\vec{x})+\tau^{\frac{3}{2}-h}\sigma_{-}(\vec{x}),
\ee
where we have separated explicitly the time depending part. The precise interpretation of the previous result depends on the value of the discriminant (\ref{h}) because depending if $m_{\phi} <\frac{3}{2}H$ or $m_{\phi} >\frac{3}{2}H$ we can provide a unitary interpretation of the conformal field theory associated, we don't enter into this discussion here which is beyond the scope of this paper and address the attention of the reader to  \cite{Strominger:2001pn}, \cite{Bousso:2001mw} and \cite{Spradlin:2001nb}. In the following, we restrict ourselves to the case $m_{\phi} <\frac{3}{2}H$.  The dominant solution for super horizon regime in this case is $\tau^{\frac{3}{2}-h}$, so we have that 
\be 
\lim_{|\tau| \rightarrow 0} {\phi} (\tau, \vec{x} ) = \tau^{\frac{3}{2}-h}\sigma_{-}(\vec{x}).
\ee 
The field $\sigma$ is the one that we need to construct the dual conformal theory for late times after horizon crossing, so, we need to determine if it behave as a conformal field by deriving its transformation under conformal transformations in $\mathbb{R}^3$, actually, as we comment before, we just need the transformation for dilatations (\ref{dilation}) and inversions  (\ref{inv})  in 3D. Before that, we need to take the super horizon limit $|\tau| \ll |\vec{x}|$ of the inversion (\ref{inv}): $\tau' = \frac{\tau}{ |\vec{x}|^2}, x'_i =  \frac{x_i}{ |\vec{x}|^2}$. Using $\phi' = \phi$, and (\ref{dilation}) we obtain
\bea
\phi'(\tau', x'_{i}) &=& \tau'^{\frac{3}{2}-h} \sigma' (\vec{x}') = \lambda^{\frac{3}{2}-h} \tau^{\frac{3}{2}-h} \sigma' (\vec{x}') \no\\
&=&  \tau^{\frac{3}{2}-h} \sigma (\vec{x}) = \phi(\tau, x_{i})
\eea
which implies:
\be \label{dilcft}
\sigma' (\vec{x}') = \lambda^{-(\frac{3}{2}-h)} \sigma (\vec{x}). 
\ee
On the other hand, for inversions in the super horizon regime  
\bal
\phi'(\tau', x'_{i}) = \tau'^{\frac{3}{2}-h} \sigma' (\vec{x}') = \frac{ \tau^{\frac{3}{2}-h} }{ |\vec{x}|^{2(\frac{3}{2}-h)} } \sigma' (\vec{x}')&\no\\
=  \tau^{\frac{3}{2}-h} \sigma (\vec{x}) =\phi(\tau, x_{i}),\ &
\eal 
which implies 
\be \label{invcft}
\sigma' (\vec{x}') =  |\vec{x}|^{2(\frac{3}{2}-h)} \sigma (\vec{x}). 
\ee
When we compare (\ref{dilcft}) and (\ref{invcft}) with  (\ref{tl}) (with $n=0$ and $d=3$) we conclude that the field $\sigma$ behaves as a conformal field with conformal weight
\be 
\Delta_{\sigma} = \frac{3}{2}{-}h= \frac{3}{2} - \sqrt{\frac{9}{4}-\frac{m_\phi^2}{H^2}} .
\ee
This is an important result given that the correlating functions of a conformal field theory can be calculated if we know the conformal weight of the primary fields. In this case, we expect that the 2 point correlating function of the dual theory of a scalar field on de Sitter space behave like
\be 
\langle {\cal O}_{\sigma} (\vec{x})  {\cal O}_{\sigma} (\vec{y}) \rangle = \frac{A}{|\vec{x} - \vec{y}|^{\Delta_{\sigma}}},
\ee 
with $A$ a constant. For an detailed analysis of the two point correlation functions in the dual theory see {\it e.g.} \cite{Strominger:2001pn} and \cite{Leblond2}.
\subsection{Single vector field.}
Now, we turn to the vector field case. We consider a massive vector field with Lagrangian 
{\small\be
S_A=\int d\tau d^3x\sqrt{-g} \left\{-\frac{1}{4}F_{\mu \nu}F^{\mu \nu}-m_A^2A_\mu A^\mu\right\},
\ee}
where $F_{\mu \nu}=\partial_\mu A_\nu-\partial_\nu A_\mu$ and $m_A$ is the mass of the vector field. This model have been studied with some detail in \cite{Ackerman:2007nb} and (allowing for non-minimal coupling) \cite{Golovnev08} and in that time, it sparked some interest in the study of inflationary models with vector fields. In de Sitter conformal coordinates (\ref{ds}) we have the action 
{\small\bal
S_A=\int d\tau d^3x \left\{-\frac{1}{2}\eta^{\mu \alpha}\eta^{\beta \nu}\partial_\nu A_\mu\left(\partial_\beta A_\alpha-\partial_\alpha A_\beta\right)\right.&\no\\
\left. -\frac{m_A^2 \eta^{\mu \nu}A_\mu A_\nu}{H^2\tau^2}\right\}.\label{svec}&
\eal}
Again, as we did in the scalar case, we will consider that $A_{\mu}$ transforms as in (\ref{tl}) in the following way  
\be
A'_\mu=(x^2)^{\Delta_A}J^\nu_\mu A_\nu,
\ee
and using inversion we get the transformation of the action with respect to the change of coordinates 
{\small \bal
S'_A=\int d\tau d^3x\frac{(x^2)^{2+2\Delta_A}}{(x^2)^4}\left\{-2\frac{(\Delta_A-1)^2}{(x^2)^2}\right.A_\mu A_\nu&\no\\
\times \left.\left(x^2\eta^{\mu \nu}-x^\mu x^\nu\right)-2\frac{(\Delta_A-1)}{(x^2)^2}x^\mu A_\alpha \eta^{\alpha \nu}\right.\times&\no\\
\left.(\partial_\mu A_\nu-\partial_\nu A_\mu)
-\frac{1}{2}\eta^{\mu \alpha}\eta^{\nu \beta}\partial_\nu A_\mu(\partial_\alpha A_\beta-\partial_\beta A_\alpha)\right.&\no\\
\left.-\frac{m_A^2 \eta^{\mu \nu}A_\mu A_\nu}{H^2\tau^2}\right\},&\no
\eal}
from which we deduce that invariance under inversions implies that $\Delta_A=1$. Again, this is not surprising since this is precisely the transformation law for a vector in four dimensional de Sitter space. \\
Now, we look if there is an associated three dimensional vector which transforms as a conformal vector field in $\mathbb{R}^3$. We begin by solving the equations of motion derived from the action (\ref{svec})
\be
\eta^{\beta \lambda}\partial_\lambda \left(\partial_\beta A_ \nu-\partial_\nu A_\beta\right)-\frac{m_A^2 A_\nu}{(H\tau)^2}=0.
\ee
The system before propagate three degrees of freedom due to the presence of the mass term, two transverse terms perpendicular to the propagation axis, and a longitudinal term for the equation of the component $A_0$. It is well known that the longitudinal mode in this model is problematic due to the presence of ghost instabilities (\cite{Himmetoglu08a}, \cite{Himmetoglu08b}), but it is also known that this problem can be solved by introducing a time dependent coupling in the Maxwell term $f(\tau) F^2$  and a time dependent mass term $m(\tau)$  (\cite{Dimopoulos09a}, \cite{Dimopoulos09vu}).
We avoid the analysis of the longitudinal mode which is not relevant for our main objectives here, we refer the reader to the references mentioned before and in the following we just focus on the transverse components that can be obtained from the divergence free spacial components ($\partial_i A_i=0$) \footnote{In general we  always can separate a vector field in a transverse and a longitudinal part: $A_{\mu} = A_{\mu}^{\bot} +  A_{\mu}^{\|}$ with $\partial^{\mu}A_{\mu}^{\bot}=0.   $}:
\be
\ddot{A}_i-\delta^{kl}\partial_k \partial_l A_i+\frac{m_A^2 A_i}{(H\tau)^2}=0.
\ee


Again, in a similar way that in the scalar case, we solve the equation for the transverse component in momentum space 
{\small\be
\tilde{A}_i(\tau,\vec{k})=\sqrt{\tau}\left(D_1 J_{p}(k\tau)+D_2 Y_{p}(k\tau)\right) \tilde{T}_i(\vec{k}),
\ee}
where $\tilde{A}_i$ is the Fourier transform of the vector field, and  
\be p=\sqrt{\frac{1}{4}-\frac{m_A^2}{H^2}}.
\ee
Going to super horizon scales at  $|k\tau| \ll 1$ and going back to coordinates space we get the asymptotic expansion 
\be
A_i(\tau,\vec{x})\approx \tau^{\frac{1}{2}+p}V_i(\vec{x})+\tau^{\frac{1}{2}-p}U_i(\vec{x}).
\ee 
For late times, again we have that the dominant term is $\tau^{\frac{1}{2} - p}$, then:
\be 
\lim_{|\tau| \rightarrow 0} A_i(\tau,\vec{x}) = \tau^{\frac{1}{2}-p} U_{i}(\vec{x}).
\ee 
And we consider the asymptotic symmetries of dilatations and inversion. The result is the same for both symmetries as in the scalar field, and then we only show the transformation law of the boundary  field $U_i$ under dilatations. Taking into account the transformation law for the vector field:
\be 
A'_i=\frac{\partial x^j}{ \partial x'^i }A_j,
\ee
which for dilatations become $A'_i = \lambda^{-1} A_i$. Then we can see that
\bea
A'_i&=&\tau'^{\frac{1}{2}-p}U'_i(\vec{x'})=\lambda^{\frac{1}{2}-p}\tau^{\frac{1}{2}-p}U'_i(\vec{x'})\no\\
&=&\lambda^{-1}A_i=\lambda^{-1}\tau^{\frac{1}{2}-p}U_i(\vec{x}),
\eea
so, $U_i$ transforms as   
\be
U'_i(\vec{x'})=\lambda^{-1-(\frac{1}{2}-p)}U_i(\vec{x}),
\ee
which, comparing with (\ref{tl}) implies that $U_i$ fulfil our expectations and behave precisely as a conformal vector field in $\mathbb{R}^3$ under the asymptotic symmetries of de Sitter group with conformal weight
\be 
\Delta_U=1 + (\frac{1}{2}-p) = \frac{3}{2}-p = \frac{3}{2}-\sqrt{\frac{1}{4}-\frac{m_A^2}{H^2}}.
\ee \\
We learned in previous sections that in the spirit of the so called dS/CFT correspondence, free fields propagating on de Sitter admit a dual conformal field theory representations in the boundary of the space which we take to be a $\mathbb{R}^3$ space placed at the super horizon limit, this is at $|\tau| \ll |\vec{x}|$. We have calculated the conformal weight of the corresponding conformal fields  correspondence which are essential for the calculation of correlation functions and to describe the statistical properties of the theory. Although the results presented in the previous sections where obtained for free fields, there are several discussions in the literature which points towards of an extension of this correspondence also in the presence of self interactions and for coupled systems involving several fields. In next section we discuss coupled system involving a scalar and a vector field which results useful for applications in inflationary cosmology and analyse in which conditions it is possible to extend the ideas explored in previous sections and if it is possible to find a conformal field theory representation of the theory aiming to calculate the properties of inflationary perturbations by using the conformal field theory machinery. 
 
\subsection{Vector-scalar coupled system.}
In this section we study a system of interacting scalar and vector fields which results useful for several applications to inflationary cosmology \cite{Ratra:1991bn,Dimopoulos09a,Watanabe10,Dimopoulos09vu,Dimopoulos10xq,Barnaby:2011vw,Sorbo:2011rz,Dimopoulos:2012av,Anber:2012du,Bartolo12,Lyth13,Shiraishi:2013kxa,Cook:2013xea,Nurmi:2013gpa,Caprini:2014mja,Chen:2014eua,Bartolo:2015dga,namba:2015gja}. The model that we  consider here is described by the action
\bea 
S_{\rm{\phi A}} &=& -\frac{1}{4}\int d^{4}x \sqrt{-g}   \left[ f_{1}(\phi) 
F^{\mu \nu}F_{\mu \nu}\right.\no\\
 &&\qquad\qquad \left. +  f_{2}(\phi)\tilde{F}^{\mu \nu}F_{\mu \nu} \right].\label{ccsv}
\eea
In the second term, $\tilde{F}^{\mu \nu}$ is the Hodge dual of the field strength $F_{\mu \nu}$ and is defined by 
\be
\tilde{F}^{\mu \nu}=\frac{1}{2\sqrt{-g}}\epsilon^{\mu \nu \alpha \beta}F_{\alpha \beta},
\ee
where $\epsilon^{\mu\nu \alpha \beta}$ is the four dimensional Levi-Civita symbol and $ f_{1}(\phi)$ and $ f_{2}(\phi)$ are coupling functions which depend only of the scalar field. This model introduces explicitly parity violation through the term $f_{2}(\phi)\tilde{F}^{\mu \nu}F_{\mu \nu}$. We assume the results obtained previously for the conformal weight of the scalar and the vector field in four dimensional de Sitter, this is $\Delta_{\phi} = 0$ and $\Delta_{A} = 1$ so that the action (\ref{ccsv}) is manifestly invariant under de Sitter group transformations. Additionally, assuming $\Delta_\phi = 0$ imply that if we assign some conformal weight $\Delta_{f_i}$ to the coupling functions $$f'_i(\phi') =\left|\det\left(\frac{\partial x'^\mu}{\partial x^\nu}\right)\right|^{-\frac{\Delta_{f_i}}{4}}   f_i(\phi), $$ we get that $\Delta_{f_i} = 0$ in order to preserve de Sitter invariance. \\
Now, we proceed as in previous sections, we solve the equations of motion and study its super horizon evolution. 
In the context that we frame the discussion we consider that the scalar field drives the inflationary dynamics and that the vector field is an auxiliary subdominant field which affects the primordial curvature perturbation and can leave an imprint on the inflationary evolution. The scalar affect the evolution of the vector field through the coupling functions $f_i(\phi)$. The equations of motion for the vector field derived from the action (\ref{ccsv}) are:
\be\label{eomg}
\nabla_{\mu}\left( f_{1} (\phi) F^{\mu \nu}  +  f_{2} (\phi) \tilde{F}^{\mu \nu}\right) = 0, 
\ee
which must be complemented with the Bianchi identity:
\be \nabla_{\mu}  \tilde{F}^{\mu \nu} = 0. \ee
 We don't solve the dynamics of the scalar field, we just assume some features of its solutions due to the fact that this field drives the inflationary expansion. Accordingly, we further assume that the inflationary dynamics homogenise the scalar perturbations,  so that we can approximate the scalar field as a time dependent function, this is $\partial_{i}\phi=0$. Accordingly, this implies that on the solutions of the inflationary scalar field $f_{1}(\phi) = f_{1} (\phi (\tau))$ and $f_{2}(\phi) = f_{2} (\phi (\tau))$,  then $\partial_{i}f_{1} = \partial_{i}f_{2}=0$. With this approximation, the temporal and spacial components of the equation (\ref{eomg})  in de Sitter conformal coordinates (\ref{ds}) become respectively
\be\label{tA}
 f_{1} \partial_{i} F_{i0} = 0,
\ee
and
{\small\be\label{iA}
(\partial_{0} f_{1}) F_{0i}  -  (\partial_{0} f_{2}) \varepsilon^{0ijk} \partial_{j}A_{k} - f_{1}\left( \partial_{j}F_{ji}  -\partial_{0}F_{0i}  \right) = 0.
\ee}    
Now, given that  this theory is manifestly gauge invariant because it only depend of the field strength $F_{\mu \nu}$ and its dual $\tilde{F}^{\mu \nu}$, we choose the Coulomb gauge and set $A_{0}=0$ and $\partial_i A_i=0$. With this choice, the temporal equation  equation (\ref{tA}) cancels and the spatial part (\ref{iA}) reduces to:
{\small \be\label{eomf1f2}
\left(\frac{\partial^{2}}{\partial \tau^{2}}    -  \nabla^{2}  + \frac{1}{f_{1}}\frac{\partial f_{1} }{\partial \tau} \frac{\partial  }{\partial \tau} +    \frac{1}{f_{1}}\frac{\partial f_{2} }{\partial \tau}  \nabla \times   \right) \vec{A}(\tau, \vec{x}) = 0.
\ee}
So far, we don't have any restriction over the form of the coupling functions $f_i$, and now we impose some conditions mainly, to preserve conformal invariance on the asymptotic region. Let's apply the dilatation $\tau' = \lambda \tau$ and $\vec{x}'= \lambda \vec{x}$  to the previous equation and demand invariance under this transformation. We further assume that the coupling functions are homogeneous functions of time, this is $f_1({\lambda \tau}) = \lambda^n f_1(\tau)$, $f_2({\lambda \tau}) = \lambda^m f_2(\tau)$, so that the equation of motion transforms as: 
\bea \left(\frac{\partial^{2}}{\partial \tau^{2}}    - \nabla^{2}  + \frac{1}{f_{1}(\tau)}\frac{\partial f_{1} (\tau)}{\partial \tau} \frac{\partial  }{\partial \tau} + \right.\no&&\\
\left.\lambda^{m-n}\frac{1}{f_{1}(\tau)}\frac{\partial f_{2}(\tau) }{\partial \tau}  \nabla \times \right)\vec{A}(\tau, \vec{x}) &=& 0. \no
\eea 
In this way we see that a necessary condition for getting scaling invariance is that the coupling functions are homogeneous functions of the same order $n=m$. This is achieved if the couplings are power law functions proportional to each other: 
\be\no
f_1(\tau) = c^2_1 (-H \tau)^n \quad {\rm and} \quad f_2(\tau) = c^2_2 (-H \tau)^n,
\ee
which implies that
\be\no
f_1(\tau) = \gamma f_2(\tau) \quad \mbox{with} \quad  \gamma = c^2_2/c^2_1 .
\ee  
In the previous expression we restored the Hubble constant for dimensional analysis.  Let us redefine $n=-2\alpha$ because the function $f_1$ must be positive due to hamiltonian stability. In order to avoid strong coupling at super horizon evolution, we assume that $\alpha<0$, then, we get:
\be \label{fsalpha}
f_1(\tau) = c^2_1 (-H \tau)^{-2\alpha} , \ f_2(\tau) = c^2_2 (-H \tau)^{-2\alpha} .
\ee
Now, let's follow the analysis in terms of the normalized canonical field $a_i$ defined as 
\be
A_i (\tau, x_i) \equiv \frac{a_i(\tau, x_i)}{\sqrt{f_1}}.
\ee
For the canonical field $a_i$, the equation of motion (\ref{eomf1f2}) is written as
\bea
\left\{\frac{\partial^{2}}{\partial \tau^{2}}    -  \nabla^{2}  +  \frac{1}{2} \left[ \frac{1}{2}  \left( \frac{\partial_{\tau} f_{1} }{f_1} \right)^2 -  \frac{\partial^2_{\tau} f_{1} }{f_1} \right]\right.&&\no\\
\left.+    \frac{1}{f_{1}}\frac{\partial f_{2} }{\partial \tau}  \nabla \times   \right\} \vec{a}(\tau, \vec{x}) = 0,&&
\eea
and using (\ref{fsalpha}) gives:
\be \label{eomxalpha}
\left(\frac{\partial^{2}}{\partial \tau^{2}}    -  \nabla^{2}  -  \frac{\alpha (\alpha + 1)}{\tau^2} - \frac{2\alpha \gamma}{\tau}    \nabla \times   \right) \vec{a}(\tau, \vec{x}) = 0.
\ee
Now, we go to Fourier space. Choosing $\vec{k} = (k, 0,0)$ and defining the transverse polarizations as $$a_{\pm} = \frac{a_y \pm i a_z}{\sqrt{2}}\,,$$ we get
\be \label{eomkalpha}
\left(\frac{\partial^{2}}{\partial \tau^{2}}    +  k^{2}  -  \frac{\alpha (\alpha + 1)}{\tau^2} \pm \frac{2\alpha \gamma k}{\tau}      \right) {\tilde{a}_{\pm}}(\tau, \vec{k}) = 0.
\ee 
The equation (\ref{eomkalpha}) can be solved analytically in terms of regular and irregular Whittaker functions $W$ and $M$:
\bal 
\tilde{a}_{\pm} (\tau, \vec{k}) = C_{1\pm}(k) W_{\alpha + 1/2}(-i \xi, 2i k\tau)&\no\\
+ C_{2\pm}(k) M_{\alpha + 1/2}(-i \xi, 2i k\tau),&
\eal
where the parameter $\xi = -\alpha \gamma$ defined in \cite{Caprini:2014mja} determine the relative size of the parity violation signals in the model. Reference \cite{Caprini:2014mja} study in full detail the features of the model with $f_1 = \gamma f_2$ and describe analytically its asymptotic behaviour. For scales in which $|k\tau| \ll \xi$ the $W$ function dominates and we will consider only its contribution. Morover, if we take $|k\tau| \ll \xi$ in (\ref{eomkalpha}) we have solutions in terms of modified Bessel functions of the first kind $\cal{I}_{\nu}$:
\bea
\tilde{a}_{\pm} (\tau, \vec{k}) \approx C_{1\pm} (k)\sqrt{-2\xi k \tau} {\cal I }_{-(1+2\alpha)}(\sqrt{-8 \xi k \tau} )&&\no\\
+ C_{2\pm} (k)\sqrt{-2\xi k \tau} {\cal I }_{(1+2\alpha)}(\sqrt{-8 \xi k \tau} ).&&\no
\eea
Which, in particular, for super horizon scales in which $|8 \xi k \tau| \ll 1$ we have 
\bea
\tilde{a}_{\pm} (\tau, \vec{k}) \approx \Big[ C_{1\pm} (k)(-2\xi k \tau)^{\alpha + 1}\Gamma(-1- 2\alpha )&&\no\\
+  C_{2\pm} (k)(-2\xi k \tau)^{-\alpha}\Gamma(1+ 2\alpha ) \Big].&&
\eea
We can write the last equation as
{\small \be\label{apmasymp}
\tilde{a}_{\pm} (\tau, \vec{k}) \approx  \tilde{u}_{\pm} (k)(-\xi H \tau)^{\alpha + 1} +  \tilde{v}_{\pm} (k)(-\xi H \tau)^{-\alpha} ,
\ee}
in which we restored $H$ and absorbed further dependence of $k, \xi, \alpha$ in the $\tilde{u}$ and $\tilde{v}$ functions. At this point we go back to coordinates space obtaining the asymptotic form:
{\small\be\label{apxmasymp}
{a}_{\pm} (\tau, \vec{x}) \approx  {u}_{\pm} (\vec{x})(-\xi H \tau)^{\alpha + 1} +  {v}_{\pm} (\vec{x})(-\xi H \tau)^{-\alpha}.
\ee}
We have succeed on separating the time and space coordinates for super horizon scales, so, we can assign a conformal boundary field in this case as in the previous cases studied. This situation is nevertheless a bit different given that the dominant term depends on the value of the exponent $\alpha$, if $\alpha>-1/2$ the term $v_{\pm}$ dominates, while, if  $\alpha<-1/2$ the term $u_{\pm}$ is the one that dominates. If $\alpha>-1/2$ we demand that\\
\be \label{bfv}
\lim_{|\tau| \rightarrow 0} {a_{\pm}} (\tau, \vec{x} ) = \tau^{-\alpha}v_{\pm}(\vec{x}),
\ee 
 while, if $\alpha<-1/2$ we take the boundary condition
\be \label{bfu}
\lim_{|\tau| \rightarrow 0} {a_{\pm}} (\tau, \vec{x} ) = \tau^{\alpha + 1}u_{\pm}(\vec{x}).
\ee   
Now, we can calculate the conformal weight of those boundary fields applying the same procedure of previous sections. We only consider the dilatation transformation, the inversion is entirely analogous. We see that the canonical field $a_{\pm}$ transforms under four dimensional dilatations $\tau' = \lambda \tau$, $x' = \lambda x$, $x'_{\pm} = \lambda x_{\pm}$ as a four dimensional vector with weight $\Delta_a =1$:
\be
a'_{\pm}=\frac{\partial x_{\pm}}{\partial x'_{\pm}} a_{\pm}=\lambda^{-\Delta_a}a_{\pm}=\lambda^{-1}a_{\pm}.
\ee
Applying dilatation over (\ref{bfv}) we have:
\bea
a'_{\pm}&=&(\tau')^{-\alpha}v'_{\pm}(\vec{x}')= \lambda^{-\alpha}(\tau)^{-\alpha}v'_{\pm}(\vec{x}')\no\\&=&\lambda^{-1}a_{\pm}=\lambda^{-1}(\tau)^{-\alpha}v_{\pm}(\vec{x}),
\eea
which implies
\be\label{udil}
v'_{\pm}(\vec{x}')= \lambda^{\alpha-1}v_{\pm}(\vec{x}),
\ee
and comparing with (\ref{tl}) with $n=1$ and $d=3$ we conclude that the boundary field $v_{\pm}$ behave as a conformal field of weight:
\be\label{cwv}
\Delta_{v} = 1-\alpha.
\ee
Following the same procedure we find that the boundary field (\ref{bfu}) behave as a conformal boundary field of weight
\be\label{cwu}
\Delta_{u} = \alpha+2.
\ee
To summarise, we have obtained that the interacting system (\ref{ccsv}) admit a conformal field theory representation at super horizon scales through the boundary vector fields $u_{\pm}, v_{\pm}$. 
A remarkable fact of this results is that they coincide with the case studied in \cite{riotto}, where they studied the model without parity violation term, this is 
\bea \label{betal}
S_{\rm{\phi A}} = -\frac{1}{4}\int d^{4}x \sqrt{-g}    f_{1}(\phi) F^{\mu \nu}F_{\mu \nu} .
\eea
Naively, we could think that both cases, the parity conserving and the parity violating models have the same conformal boundary field associated and that the statistical features of the theories given by the correlating functions are the same. Nevertheless, there is a crucial difference implied by the fact that the super horizon boundary symmetry group allow the spacial reflection $\vec{x}' = - \vec{x}$ as an element of the symmetry group. In this case, the correlating functions allow for antisymmetric terms in its structure, for instance, the power spectrum of a boundary field would be
{\small\be
\langle u_i (\vec{k})  u_j (-\vec{k}) \rangle = (\beta_1(k)\delta_{ij} - \beta_2(k) \hat{k}_i \hat{k}_j + {\beta_3(k)\epsilon_{ija}k_a })  
\ee}
where the form of the $\beta_i$ functions can be determined by using the asymptotic symmetries of the theory \cite{riotto}. It is important to see that the functions $\beta_2(k)$ and  $\beta_3(k)$ are related to anisotropic and to parity violating effects respectively. In the parity conserving case ($\beta_3=0$), 
the above equation can be written as 
{\small\be
\langle u_i (\vec{k})  u_j (-\vec{k}) \rangle = \beta_1(k)(\delta_{ij} - \hat{k}_i \hat{k}_j ) \;. 
\ee}
If we assume that the vector field points along the direction of some unit vector $\hat{\bf n}$ and the wave vector ${\bf k}$ points along the unit vector $\hat{\bf k}$, the power spectrum can be written as \cite{Ackerman:2007nb}: 
\be
\beta({\bf k})=\beta(k)(1+g(k)(\hat{k}\cdot \hat{n})^2)\;,
\ee
where $\beta(k)$ is the isotropic power spectrum  and $g(k)$ is a function measuring the amount of statistical anisotropy. If  $g(k)$ is scale invariant, recent data analysis gives an upper bound on that: $g\leq 10^{-2}$ \cite{Kim:2013gka,Ade:2015hxq,Ade:2015lrj}.  

A further development of this ideas on the calculation of the correlation functions using conformal field theory techniques exploiting the asymptotic symmetries as presented in \cite{riotto} for the case of parity violating theories have enough interests and we expect to pursue this possibility in future work.   

\section{Conclusions}
In this paper we have studied, in a comprehensive way, the symmetry properties of the scalar and vector fields solutions on de Sitter space, we study their asymptotic limit at super horizon scales and we found that, in this limit, free fields behave as conformal fields on a constant time surface. We study the free scalar and the free vector fields for pedagogical purposes, to illustrate the technique and then study an inflationary model with interacting scalar and vector fields.
The coupled system is studied 
because it is interesting for its cosmological implications, in particular this kind of model can explain some observable CMB anomalies such as broken of inversion and rotational invariance.  The super horizon scales are important in our context because inflationary expansion freeze the perturbations at this scales and the correlation functions, evaluated at this scales, carry information from the inflationary dynamics. In this sense, the use of the asymptotic conformal symmetries of de Sitter space offers a powerful technique to uncover the structure of the correlations, which relies only of the conformal weight of the boundary asymptotic fields. Regarding the specific scalar-vector model studied, we can mention two important aspects, the first one is that the  coupling functions  $ f_{1}(\phi)$ and $ f_{2}(\phi)$ in Eq. (\ref{ccsv}) fulfill the condition $f_{2}(\phi)/f_{1}(\phi)=constant$ and the second one is that if we assume that the functions are a power law on $H\tau$ (which is a good choice because we are looking for an homogeneous universe), the conformal weight of the vector fields on super horizon scales depend only of that power (see Eqs. (\ref{cwv}) and (\ref{cwu})). This result is according with the one obtained by \cite{riotto}, who studied the parity conserving case with 
 $f_{2}(\phi)=0$. Our next step is to use the conformal weight of the boundary vector field to calculate the shape of the correlation functions following the procedure given by \cite{riotto} and identify the signals of parity violation.
\section{Acknowledgements} 
This work was partially supported by COLCIENCIAS grants numbers 110656933958 RC 0384-2013 and 123365843539 RC FP44842-081-2014 and by COLCIENCIAS-ECOS NORD grant number RC 0899-2012. C.A.V.-T. acknowledges financial support from Vicerrector\'ia de Investigaciones (Univalle) grant number 7924. JPBA thanks Universidad del Valle for its warm hospitality during several stages of this project.




\bibliographystyle{unsrt} 
\bibliography{Biblio} 

\end{multicols}

\end{document}